\documentclass[11pt]{article}
\pdfoutput=1
\usepackage{jheppub}
\usepackage{graphicx}
\usepackage{amsfonts,amssymb,amsmath}
\usepackage{mathrsfs}
\usepackage[hang]{subfigure}
 \usepackage[utf8]{inputenc}
\usepackage[a4paper,top=1.89in, bottom=0in, left=2in, right=0in]{geometry}
\usepackage{xcolor}

\usepackage{color}

\newcommand{\be}{\begin{equation}}
\newcommand{\ee}{\end{equation}}
\newcommand{\ba}{\begin{eqnarray}}
\newcommand{\ea}{\end{eqnarray}}

\title{Thermodynamics of Gauss-Bonnet-de Sitter Black Holes}
\author[1]{Sumarna Haroon}
\author[2]{Robie A. Hennigar}
\author[3,4]{Robert B. Mann}
\author[3,4]{Fil Simovic}

\affiliation[1]{Department of Mathematics, School of Natural Sciences,
	National University of Sciences and Technology, H-12, Islamabad, Pakistan}
\affiliation[2]{Memorial University of Newfoundland,\\
	230 Elizabeth Ave, St. John's, Newfoundland A1C 5S7, Canada}
\affiliation[3]{Perimeter Institute for Theoretical Physics,\\
	31 Caroline St. N.,Waterloo, Ontario N2L 2Y5, Canada}
\affiliation[4]{Department of Physics and Astronomy, University of Waterloo,\\
	Waterloo, Ontario N2L 3G1, Canada}
\emailAdd{sumarna.haroon@sns.nust.edu.pk}
\emailAdd{rhennigar@mun.ca}
\emailAdd{rbmann@uwaterloo.ca}
\emailAdd{fil.simovic@gmail.com}

\abstract{We investigate the thermodynamics of Gauss-Bonnet black holes in asymptotically de Sitter spacetimes embedded in an isothermal cavity, via a Euclidean action approach. We consider both charged and uncharged black holes, working in the extended phase space where the cosmological constant is treated as a thermodynamic pressure. We examine the phase structure of these black holes through their free energy. In the uncharged case, we find both Hawking-Page and small-to-large black hole phase transitions, whose character depends on the sign of the Gauss-Bonnet coupling. In the charged case, we demonstrate the presence of a {\it swallowtube}, signaling a compact region in phase space where a small-to-large black hole transition occurs.}


\begin{document}
\maketitle

\section{Introduction}

Nearly half a century after Hawking famously discovered that black holes radiate, the thermodynamics of black holes continues to serve as a lamppost guiding research into quantum gravity. Like ordinary substances, black holes possess temperature, entropy, and other thermodynamic potentials with variations between equilibrium configurations being captured by the first law of thermodynamics. Remarkably, this similarity with ordinary substances extends also to phase transitions --- Hawking and Page demonstrated the existence of a first-order phase transition between thermal radiation and a large anti-de Sitter black hole~\cite{Hawking:1982dh}.

The last decade has seen a resurgence of interest in the thermodynamics --- and in particular, the phase structure --- of black holes in the presence of a cosmological constant. This interest is due in large part to the observations of Kastor, Ray, and Traschen that a new thermodynamic potential enters into the derivation of the Smarr formula  in the presence of a non-zero cosmological constant: the thermodynamic volume~\cite{Kastor:2009wy}. The thermodynamic volume can be understood as the  quantity conjugate to the cosmological constant, interpreted in this context as a pressure, and appears as such in the first law of thermodynamics when variations of the cosmological constant are included. There has since been considerable development of these ideas including a proposed bound on the black hole entropy in terms of the thermodynamic volume~\cite{CveticEtal:2010}, the notion of holographic heat engines~\cite{Johnson:2014yja}, extensions to include acceleration, going beyond black holes to spacetimes with non-trivial topology~\cite{Appels:2016uha, Bordo:2019tyh, Andrews:2019hvq}, and connections with holography~\cite{Karch:2015rpa, Couch:2016exn, Sinamuli:2017rhp, Andrews:2019hvq}. Perhaps most actively investigated has been the subject of black hole phase transitions where examples of van der Waals behaviour~\cite{Kubiznak:2012wp}, triple points~\cite{Altamirano:2013uqa} (like that of water), (multiple) re-entrant phase transitions~\cite{Altamirano:2013ane, Frassino:2014pha} (like those occurring in certain gels), and even lambda transitions (like those marking the onset of superfluidity)~\cite{Hennigar:2016xwd} have been observed. We refer the reader to the review~\cite{Kubiznak:2016qmn} where a number of these developments are summarized.

Most of these investigations pertain to anti-de Sitter black holes, while the case of de Sitter black holes has seen comparatively little development~\cite{Dolan:2013ft, Kubiznak:2015bya, Mbarek:2016mep}. Notwithstanding the inherent difficulties, there are good reasons for exploring these ideas in the de Sitter realm. Not only could such results be relevant theoretically within the dS/CFT correspondence~\cite{Strominger:2001pn}, but it is widely accepted that our own universe possesses a positive cosmological constant. More pragmatically, it is of interest to understand how general a feature the phase structure of anti-de Sitter black holes is: do the same types of phase transitions manifest for de Sitter black holes, or are there new examples? 

The study of thermodynamics of de Sitter black holes faces an immediate problem due to the presence of the cosmological horizon: with a temperature generically different from the black hole horizon, the system is manifestly out of equilibrium. However, there are ways in which progress can still be made.\footnote{See also~\cite{Dinsmore:2019elr, Johnson:2019ayc} for other recent developments on this subject.} One method is to fix various parameters of the system to set the two horizon temperatures to be equal --- see, {\it e.g.},~\cite{Mbarek:2018bau}. A second method, and the one we will be concerned with here, involves confining the black hole within a perfectly reflecting cavity, a method originally developed in the asymptotically flat setting~\cite{York:1986it, Braden:1990hw}. The cavity approach was first applied to de Sitter black holes in~\cite{Carlip:2003ne}, though that analysis did not consider aspects of the extended thermodynamics. Recently generalizations of this approach to include considerations of the extended thermodynamics have been used to uncover examples of \textit{compact} van der Waals-like transitions for charged de Sitter black holes~\cite{Simovic:2018tdy} and re-entrant phase transitions for de Sitter black holes with nonlinear electrodynamics~\cite{Simovic:2019zgb}.

Here we further pursue the cavity approach and explore the role of higher-curvature corrections to de Sitter black hole thermodynamics. Higher-curvature corrections are ubiquitous in approaches to quantum gravity where they arise as quantum corrections to the Einstein-Hilbert action. Here we consider Lovelock theory~\cite{Lovelock:1971yv}, and in particular Gauss-Bonnet gravity, which is the simplest member of the Lovelock class beyond Einstein gravity. Lovelock gravity is in many ways a natural generalization of Einstein gravity to higher dimensions, maintaining the property of having second-order field equations for all backgrounds. Moreover, the boundary terms for Lovelock theory have long been known~\cite{Myers:1987yn, Teitelboim:1987zz, Davis:2002gn}, which is advantageous for the present study. In the realm of AdS black holes, higher-curvature theories have resulted in a number of interesting observations~\cite{Wei:2012ui, Cai:2013qga, Xu:2013zea, Mo:2014qsa, Wei:2014hba, Mo:2014mba, Zou:2013owa, Belhaj:2014tga, Xu:2014kwa, Frassino:2014pha, Dolan:2014vba, Sherkatghanad:2014hda, Hendi:2015cka, Hendi:2015oqa, Hennigar:2015esa, Hendi:2015psa, Nie:2015zia, Hendi:2015pda, Hendi:2015soe, Zeng:2016aly, Hennigar:2016gkm, EricksonRobie, Hennigar:2016xwd, Cvetic:2010jb, Hennigar:2014cfa, Johnson:2014yja, Karch:2015rpa, Caceres:2015vsa, Dolan:2016jjc, Sinamuli:2017rhp, Li:2017wbi, Dehyadegari:2018pkb, Hendi:2018xuy}, most notably being (multiple) re-entrant phase transitions, triple points, and $\lambda$-type superfluid transitions~\cite{Frassino:2014pha,Hennigar:2016xwd}. One of our goals here is to explore to what extent these interesting features carry over   to the de Sitter case. Additionally, older~\cite{Louko:1996jd} and more recent~\cite{Wang:2019urm} works have considered the thermodynamics of asymptotically flat Gauss-Bonnet black holes in cavities --- our work can be considered the natural generalization of these setups to de Sitter space.

Our paper is organized as follows: In Section 2, we present the Gauss-Bonnet theory of gravity, defining the action, metric function, and relevant boundary terms. In Section 3, we consider uncharged black holes. The on-shell action is evaluated and all relevant thermodynamic quantities are calculated. We use the first law to derive the conjugate variables. We also construct the free energy of the spacetime and study its phase structure. In Section 4, we repeat this analysis for charged black holes. We conclude with a summary of the results in Section 5.

\section{Gauss-Bonnet Gravity}

Our aim is to study the phase structure of de Sitter black holes including higher-curvature corrections to the action. The phase structure is obtained via an analysis of the free energy which can in turn be obtained from the Euclidean on-shell action for general theories of gravity. To leading order in the semi-classical approximation, the on-shell Euclidean action, $I_{\rm E}$ is directly related to the free energy by
\be 
F = - T \log Z = T I_{\rm E} \, .
\ee
As in~\cite{York:1986it, Braden:1990hw, Carlip:2003ne, Simovic:2018tdy, Simovic:2019zgb}, we shall impose that the black hole resides in a perfectly reflecting cavity which necessitates a Dirichlet boundary condition at the location of the cavity. The temperature of the cavity will be held fixed and will generically be different than the temperature associated with the cosmological horizon. As a prototypical model for higher-curvature corrections we use Gauss-Bonnet gravity which has the following (Euclidean) action:
\begin{align}\label{action!}
I_{\rm E} =&  -\int_{\cal M} d^D x \sqrt{g}  \left\{\frac{1}{16 \pi G} \left[R-2 \Lambda  + \frac{\lambda_{\rm GB} \mathcal{X}_4}{(D-3)(D-4)} \right] - \frac{1}{4} F_{\mu\nu}F^{\mu\nu} \right\}
\nonumber\\
&- \frac{1}{8 \pi G} \int_{\partial \mathcal{M}} d^{D-1} x \sqrt{\gamma} \left[K + \frac{2 \lambda_{\rm GB}}{(D-3)(D-4)} \left[\mathcal{J} - 2 \mathcal{G}_{ij} K^{ij} \right] \right] 
\nonumber\\
&- \int_{\partial \mathcal{M}} d^{D-1} x \sqrt{\gamma} F^{ij} n_i A_j\, ,
\end{align}
where we have included a Maxwell field in addition to the gravitational terms. Here, $R$ is the Ricci scalar, $\Lambda$ is the cosmological constant, $\lambda_{GB}$ is the Gauss-Bonnet coupling (which has units of inverse length squared), $\mathcal{X}_4$ is the Euler density, and $F_{\mu\nu}$ is the electromagnetic field strength tensor. The terms appearing in the first line are the usual bulk terms from which the equations of motion are derived. The boundary terms ensuring a well-posed Dirichlet problem appear in the second line. Finally, the third line contains the relevant boundary term for the Maxwell field to ensure that the system is in the fixed charge ensemble. Here $g_{\mu\nu}$ is the full spacetime metric, while $\gamma_{ij}$ is the induced metric on the boundary. The vector $n_\mu$ is the outward pointing normal to the constant $r$ hypersurface. For a constant $r$ surface in a (Euclidean) spherically symmetric geometry, this boundary metric takes the form
\be 
\gamma_{ij} dx^i dx^j = f(r) dt_{\rm E}^2 + r^2 d\Sigma_{k, D-2} \, .
\ee
The object that appears in the boundary action --- $\sqrt{\gamma}$ --- is the square root of the determinant of this metric. $\mathcal{J}$ is the trace of the boundary tensor,
\be
\mathcal{J}_{ij} = \frac{1}{3} \left(2 K K_{ik}K^k_j + K_{kl}K^{kl}K_{ij} - 2 K_{ik}K^{kl}K_{lj} - K^2 K_{ij} \right) \, ,
\ee
with $K_{ij}$ the extrinsic curvature and $K = h^{ij}K_{ij}$ its trace, $\mathcal{G}_{ij}$ is the Einstein tensor computed for the boundary metric $\gamma_{ij}$, and the Euler density $\mathcal{X}_4$ is given by
\be
\mathcal{X}_4 = R_{\mu\nu\sigma\rho}R^{\mu\nu\sigma\rho} - 4 R_{\mu\nu} R^{\mu\nu} + R^2 \, .
\ee
For the (Lorentzian) metric and gauge field we have that\footnote{Note that the one-form $dt$ diverges on the horizon. We have chosen a gauge for $A$ such that this divergence does not lead to an ill-defined gauge field on the horizon.}
\be
ds^2 = - f(r) dt^2 + \frac{dr^2}{f(r)} + r^2 d\Omega^2 \, ,
\ee
\be
A_{\mu} = -\frac{1}{2\sqrt{2\pi G}}\sqrt{\frac{D-2}{D-3}} \left[\frac{q}{r^{D-3}}-  \frac{q}{r_+^{D-3}}\right]dt =\left[\phi(r)- \phi(r_+) \right] dt\, ,
\ee
and the field equations reduce to a polynomial equation that determines the metric function $f(r)$,
\begin{equation}\label{BHeq}
h\left(\frac{(f(r)-1)}{r^2}\right)=\frac{\omega_{D-3}}{r^{D-1}}-\frac{q^2 }{r^{2(D-2)}}\, ,
\end{equation}
with $h(x)$ given by the polynomial function
\begin{equation}
h(x)=-\frac{2\Lambda}{(D-1)(D-2)}-x+\lambda_{\rm GB} x^2 \, .
\end{equation}
In these expressions, $q$ and $\omega$ are two integration constants that are related to the mass $M$ and charge $Q$ of the black hole according to
\begin{eqnarray}
\omega_{D-3}&=&\frac{16\pi G M}{(D-2)\Omega_{D-2}}\, ,\\
q&=&\frac{ Q}{\Omega_{D-2}}\sqrt{\frac{8 \pi G}{(D-2)(D-3)}}\, .
\end{eqnarray}
Note that when $\lambda_{\rm GB} = 0$ we get
\be
f(r) = 1 - \frac{\omega_{D-3}}{r^{D-3}} + \frac{q^2}{r^{2(D-3)}} - \frac{2 \Lambda r^2}{(D-1)(D-2)} \, ,
\ee
which is the ordinary charged (A)dS black hole solution in Einstein gravity. However, here we will be interested in the case where $\lambda_{\rm GB} \neq 0$ and will work with $D \ge 5$. In this case the metric function is the solution of a quadratic equation, and we pick the root that has a smooth limit as $\lambda_{\rm GB} \to 0$.

\subsection{Calculating the on-shell action}

In this section we will compute the on-shell action. Working quite generally, we consider a metric of the following (Euclidean) form:
\be
ds^2 = f(r) dt^2 + \frac{dr^2}{f(r)} + r^2 d\Sigma_{k, D-2}
\ee
where $d\Sigma$ is the line element on a space of constant curvature with $k \in \{-1,0,1 \}$ denoting negative, zero, and positive curvature. The specific form of the line element can be found, for example, in Eq.~(4) of \cite{Hennigar:2015esa}. Using the methods of~\cite{Deser:2005pc}, it is quite straight-forward to perform a direct computation of the on-shell action in any spacetime dimension. The following terms contribute to the bulk action:
\begin{align}
R &= -\left[f'' + \frac{2(D-2) f'}{r} - \frac{(D-2)(D-3)(k-f)}{r^2} \right]\, ,\nonumber\\
R^2 &= \left[f'' + \frac{2(D-2) f'}{r} - \frac{(D-2)(D-3)(k-f)}{r^2} \right]^2 \, ,
\nonumber\\
R_{\mu\nu}R^{\mu\nu} &= \frac{1}{2} \left(f'' + \frac{(D-2) f'}{r} \right)^2 + (D-2) \left(-\frac{f'}{r} + \frac{(D-3)(k-f)}{r^2} \right)^2 \, ,
\nonumber\\
R_{\mu\nu\sigma\rho}R^{\mu\nu\sigma\rho} &= (f'')^2 + 2(D-2) \left(\frac{f'}{r} \right)^2 + 2(D-2)(D-3)	\left(\frac{k-f}{r^2} \right)^2 \, .
\end{align}
Some algebra allows us to recognize that the bulk gravitational action is a total derivative:
\be\label{BulkAct}
I_{\rm E}^{\rm bulk} = - \frac{\Omega_{D-2} \beta }{16 \pi G} \frac{d}{dr} \left[ - \frac{2(D-2) q^2}{r^{D-3}} + (D-2) \omega_{D-3} - r^{D-2} f' \left(1- \frac{2 (D-2)  \lambda_{\rm GB}}{D-4} \frac{(f-k)}{r^2} \right) \right] \, ,
\ee
In producing this expression we have made use of the field equations. Specifically, we have replaced the appearance of a $r^{D-1} h$ term with the corresponding factors of $\omega$ and $q$. Note that, in the above, $\beta$ is the periodicity of the Euclidean time enforced by demanding no conical singularities at the zero of $f$.

Let us now focus on computing the boundary term. The Euclidean solution is a smooth manifold at the horizon with topology $\mathbb{R}^2 \times \mathbb{S}^{D-2}$. We therefore consider a boundary term only at the location of the cavity. To compute the boundary term, we use the convenient notation of~\cite{Deser:2005pc} which introduces the orthonormal projectors 
\be
\tau_\mu^{\ \nu} = \delta_\mu^{\ t} \delta^{\ \nu}_{t}  \qquad  \rho_\mu{}^{\nu} = \delta_\mu^{\ r} \delta^{\ \nu}_{r}
\qquad \sigma_\mu^{\ \nu} =  \sum_{i=1}^{D-2} \delta_\mu^{\ i}\delta^{\ \nu}_{i}
\ee
to decompose the curvature into temporal, radial, and angular parts (in the last term the sum extends over the angular directions). These orthogonal projectors satisfy the following relations: 
\be
\tau_{\mu}^{\ \lambda} \tau_{\lambda}^{\ \nu} = \tau_{\mu}^{\ \nu} \, , \quad \rho_{\mu}^{\ \lambda} \rho_{\lambda}^{\ \nu} = \rho_{\mu}^{\ \nu} \,,\quad \sigma_{\mu}^{\ \lambda} \sigma_{\lambda}^{\ \nu} = \sigma_{\mu}^{\ \nu}\, ,
\ee
and
\be
\tau_{\mu}^{\ \nu} \tau_{\nu}^{\ \mu} = \rho_{\mu}^{\ \nu} \rho_{\nu}^{\ \mu} = 1  \, , \quad \sigma_{\mu}^{\ \nu} \sigma_{\nu}^{\ \mu} = D - 2\, .
\ee
The boundary term is composed of various traces and contractions of the extrinsic curvature tensor $K_{ij}$. For a constant $r$ surface, the extrinsic curvature computed for the outward-pointing unit normal vector is
\be
K_i^j =  \frac{f'}{2 \sqrt{f}} \tau_i^j + \frac{\sqrt{f}}{r} \sigma_i^j \, ,
\ee
and the curvature tensor of the boundary is
\be
{\cal R}_{ij}{}^{kl} = \frac{2 k }{r^2} \sigma_{[i}^k \sigma_{j]}^l \, ,
\ee
where $k$ characterizes the curvature of the constant time slices of the boundary, as mentioned above. From the Riemann tensor we compute the Ricci tensor and Ricci scalar of the boundary to be
\be
{\cal R}_i^j = \frac{(D-3)k}{r^2} \sigma_i^j \, , \quad {\cal R} = \frac{(D-3)(D-2)k}{r^2} \, .
\ee
We then note that the Einstein tensor of the boundary geometry is just given by
\be
{\cal G}_i^j = {\cal R}_i^j - \frac{1}{2} \delta_i^j {\cal R} = \frac{(D-3) k }{r^2} \left[\sigma_i^j - \frac{D-2}{2} \delta_i^j \right] \, .
\ee
Using these results, some simple manipulations yield the following results
\begin{align}
K  &=  \frac{1}{2 \sqrt{f}} \left[f' + \frac{2 (D-2)}{r} f \right] \, ,
\nonumber\\
K_i^jK_j^i &= \left(\frac{f'}{2 \sqrt{f}} \right)^2 + (D-2) \left(\frac{\sqrt{f}}{r} \right)^2 \, ,
\nonumber\\
K_i^jK_j^lK_l^i &= \left(\frac{f'}{2 \sqrt{f}} \right)^3 + (D-2) \left(\frac{\sqrt{f}}{r} \right)^3 \, .
\end{align}
Putting these together we obtain
\begin{align}
{\cal J} &= \frac{1}{3} \left\{\frac{3}{2 \sqrt{f}} \left[f' + \frac{2 (D-2)}{r} f \right] \left[\left(\frac{f'}{2 \sqrt{f}} \right)^2 + (D-2) \left(\frac{\sqrt{f}}{r} \right)^2 \right]   - 2 \left[\left(\frac{f'}{2 \sqrt{f}} \right)^3 + (D-2) \left(\frac{\sqrt{f}}{r} \right)^3 \right]  \right.
\nonumber\\
& \left.-  \left(\frac{1}{2 \sqrt{f}} \left[f' + \frac{2 (D-2)}{r} f \right] \right)^3 \right\}
\end{align}
and
\be
{\cal G}_{ij}K^{ij} = - \frac{(D-2)(D-3)k}{2 r^2} \left[(D-4) \left(\frac{\sqrt{f}}{r} \right)  +  \frac{f'}{2 \sqrt{f}} \right] \, ,
\ee
where primes denote derivatives with respect to $r$. With this in place, we can calculate explicitly the on-shell action for the Gauss-Bonnet black hole.

In the following sections we will consider the uncharged case and the charged cases separately. We will work in the fixed charge ensemble, which requires the addition of the Maxwell boundary term appearing in the last line of Eq.~\eqref{action!}. Noting that the outward pointing normal one-form is
\be 
n_\mu dx^\mu = \frac{dr}{\sqrt{f(r)}}
\ee
and taking care to work with the Euclideanized gauge potential\footnote{In other words, requiring that $q \to i q$ so that $A_t dt = A_{t_E} dt_E$ --- see, e.g.,~\cite{Braden:1990hw} for additional details.} it can easily be shown that
\be 
\int_{\partial \mathcal{M}}\sqrt{\gamma} F^{ij}n_i A_j = \frac{(D-2)\beta \Omega_{D-2}}{8 \pi G} \left[\frac{q^2}{r_c^{D-3}} - \frac{q^2}{r_+^{D-3}} \right]  = \frac{(D-2)\beta \Omega_{D-2}}{8 \pi G} \left[\frac{q^2}{r^{D-3}} \right]_{r=r_+}^{r=r_c}\, .
\ee
From the expression for the bulk action in \eqref{BulkAct} it is then obvious that when this term is subtracted from the bulk action the explicit charge dependence completely drops out. In the fixed charge ensemble, the charge appears in the action only through its appearance in $f(r)$.


\section{Uncharged Gauss-Bonnet Black Holes}

We begin with a study of the thermodynamic properties of $D$-dimensional uncharged Gauss-Bonnet-de Sitter black holes. Computing the full on-shell Euclidean action is straightforward since the bulk action is a total derivative. Upon integration it gives two contributions: one at the horizon $r_+$, and one at the location of the cavity, $r_c$. We will also specialize to the case $k=1$, so that the transverse sections are spheres. The boundary term contributes only at the cavity. Performing the action calculation, followed by some simplification, we arrive at the following general result:
\begin{align}
I_{\rm E} =\  &\frac{(D-2)\Omega_{D-2} \beta f(r_c) r_c^{D-5}}{24 \pi G} \left( -3 r_c^2 + 2 \lambda_{\rm GB}(f(r_c) - 3)\right) \nonumber\\
&- \frac{\Omega_{D-2} r_+^{D-2}}{4 G} \left[1 + \frac{2(D-2)\lambda_{\rm GB}}{(D-4) r_+^2} \right]  \, .
\end{align}
Here $\beta$ is the periodicity of the Euclidean time, which is  redshifted to a value $ \sqrt{f(r_c)} \beta $ at the cavity.  We wish to physically fix the temperature of the boundary to be this value, so that $\beta_c = \sqrt{f(r_c)} \beta $, thereby ensuring thermodynamic equilibrium within the cavity.

The answer above is not quite complete. It is customary to normalize the action such that flat, empty spacetime has zero action and energy. To achieve this, we must subtract from the action the boundary term evaluated for an identical cavity in flat spacetime. This subtraction term has the form
\begin{align} 
I_0 = - \frac{(D-2) \Omega_{D-2} \beta_c r_c^{D-5}}{8 \pi G} \left[r_c^2 + \frac{4}{3} \lambda_{\rm GB} \right] \, ,
\end{align}
and the complete action is then $I_{\rm E} - I_0$. In the present work, especially in the context of uncharged black holes, we will be interested in comparing the free energy of the black hole solutions with the free energy of an identical cavity filled with radiation. The free energy of the latter configuration is obtained from setting the metric function $f(r)$ to be the one for a pure de Sitter solution in $I_{\rm E} - I_0$.\footnote{Note that, in the case of flat asymptotics, the action and energy for an empty cavity are set to zero simply by subtracting a boundary term for an identical cavity embedded in flat spacetime. However, in the dS case, a subtraction of the boundary term for an identical cavity in pure dS will not accomplish this --- the reason is that when the cosmological constant is nonzero the bulk action contributes also to the total action.} For convenience, we then consider the difference between these two actions which is equal to
\begin{align} 
I_* \equiv \Delta (I_{\rm E} + I_0) &= \frac{(D-2)\Omega_{D-2} \beta_c  r_c^{D-5}}{24 \pi G} \left[\sqrt{f(r_c)} \left( -3 r_c^2 + 2 \lambda_{\rm GB}(f(r_c) - 3) \right) \right.
\nonumber\\
&\left.- \sqrt{f_0(r_c)}\left( -3 r_c^2 + 2 \lambda_{\rm GB}(f_0(r_c) - 3) \right) \right] - \frac{\Omega_{D-2} r_+^{D-2}}{4 G} \left[1 + \frac{2(D-2)\lambda_{\rm GB}}{(D-4) r_+^2} \right]  \, ,
\end{align}
where $f_0(r)$ denotes the metric function with the mass parameter set to zero and $f(r)$ is the full solution for the physical Gauss-Bonnet-de Sitter black hole
\be
f(r) = 1+ \frac{r^2}{2 \lambda_{\rm GB}} - \frac{r^{2 - D/2} \sqrt{r^D+4\,\lambda_{\rm GB} \big(r\, \omega_{D-3} + \,r^D/L^2\big)}}{2 \lambda_{\rm GB}}  \,
\ee
where we have defined 
\be 
\Lambda \equiv \frac{(D-1)(D-2)}{2 L^2}  
\ee
in the above. It is this branch that reduces appropriately to the Einstein gravity solution when the Gauss-Bonnet coupling is turned off. However note also that when the mass parameter is set to zero  the metric function becomes
\be\label{f0}
f_0(r) = 1 + \left(\frac{1 - \sqrt{1 + 4 \lambda_{\rm GB} /L^2} }{2 \lambda_{\rm GB} } \right)r^2
\ee
which corresponds to the pure dS vacuum of the theory. Note that the higher-curvature corrections `renormalize' the cosmological constant. The asymptotics will be sensible provided that $\lambda_{\rm GB}/L^2 > -1/4$. This will be the case in the bulk of this work where we focus primarily on positive coupling. If this bound is violated then the solution will terminate at some value of $r$ and will not extend all the way $r \to \infty$.


From the above it is now possible to compute the entropy and energy of the solutions in the standard way. We find
\begin{align}\label{eands}
E =\ & \frac{\partial I_*}{\partial \beta_c} = \frac{(D-2)\Omega_{D-2} r_c^{D-5}}{24 \pi G} \left[ \sqrt{f(r_c)} \left(-3 r_c^2 + 2 \lambda_{\rm GB} (f(r_c) - 3) \right) \right.
\nonumber\\
&\left.\qquad\ \ \,- \sqrt{f_0(r_c)} \left(-3 r_c^2 + 2 \lambda_{\rm GB} (f_0(r_c) - 3) \right) \right]\, ,
\nonumber\\
S =\ & \beta_c E - I_* = \frac{\Omega_{D-2} r_+^{D-2}}{4 G} \left[1 + \frac{2(D-2)\lambda_{\rm GB}}{(D-4) r_+^2} \right] 	\, ,
\end{align}
where we note that, since $I_*$ is the difference in actions of the black holes and the cavity filled with thermal gas, $E$ here corresponds to difference in energies between those solutions. Note also that since we are working on-shell, the computation of these quantities needs to account for the fact that $\omega$ is not independent of $\beta$. The entropy here is exactly the Iyer-Wald entropy computed for Gauss-Bonnet black holes with spherical horizons. The energy has received ``self energy'' corrections due to the presence of the cavity.\footnote{In the case where $\Lambda = 0$ we can easily see that, in the limit $r_c \to \infty$ \be
\lim_{r_c \to \infty} E = \frac{(D-2) \Omega_{D-2} \omega_{D-3}}{16 \pi G} = M \nonumber
\ee
which matches precisely our expectations.}
Note that the energy implicitly depends on $r_+$ due to the appearance of the mass parameter $\omega_{D-3}$ in the metric function.

The temperature $T=\beta^{-1}$ is obtained by demanding that the variation of $I_*$ with respect to $r_+$ vanish. This is accomplished by first rewriting $f(r_c)$ in terms of $r_+$, by isolating for $M$ in $f(r=r_+)=0$ and substituting back into $f$. We find that
\be\label{temp}
\frac{\partial I_*}{\partial r_+}=0\quad\rightarrow\quad \beta^{-1}=T= \frac{f'(r)\big|_{r_+}}{4 \pi \sqrt{f(r_c)}} \, ,
\ee
where again the prime indicates a derivative with respect to $r$. This result is consistent with our expectation that the temperature required for equilibrium in the cavity should coincide with the redshifted Hawking temperature at the location of the cavity. From here on, we work in natural units where $G=c=1$. In this way, all lengths are measured in units of $l_p$.

\subsection{The first law}

The (extended) first law of thermodynamics for uncharged Gauss-Bonnet black holes reads:
\be\label{firstlaw}
dE=TdS+VdP+\sigma dA+\Phi_{GB}\lambda_{GB}\, .
\ee
Here, the pressure--volume term $VdP$ appears since we are considering variations in the cosmological constant in the extended phase space, where the pressure is related to $\Lambda$ through
\be
P=-\dfrac{\Lambda}{8\pi}=\dfrac{(D-1)(D-2)}{16\pi L^2}\ ,
\ee
and $V$ is the thermodynamic volume of the system, which in general differs from the geometric volume of the black hole. Additionally, a work term $\sigma dA$ associated with changes in the cavity area $A$ is present, where $\sigma$ is interpreted as the surface tension of the cavity. Finally, the $\Phi_{GB}\lambda_{GB}$ term must be included to account for variations in the Gauss-Bonnet coupling. Having determined the energy $E$, temperature $T$, and entropy $S$, we can determine what the conjugate variables $\{V,\lambda,\Phi_{GB}\}$ must be for the first law to hold. Expressions for the conjugate variables are lengthy and can be found in Appendix A. The first law \eqref{firstlaw} thus holds by construction.
\\

One can also show that these variables satisfy the Smarr relation, which in this case reads:
\be
(D-3)E=(D-2)TS-2PV+2\Phi_{GB}\lambda_{GB}
\ee
This relation is broadly applicable as it holds for both asymptotically AdS and dS spacetimes, is valid in any dimension, and is also satisfied by more exotic objects like black rings and black branes~\cite{Kastor:2009wy}. 

\begin{figure}[h]
	\centering
	\includegraphics[width=0.4\textwidth]{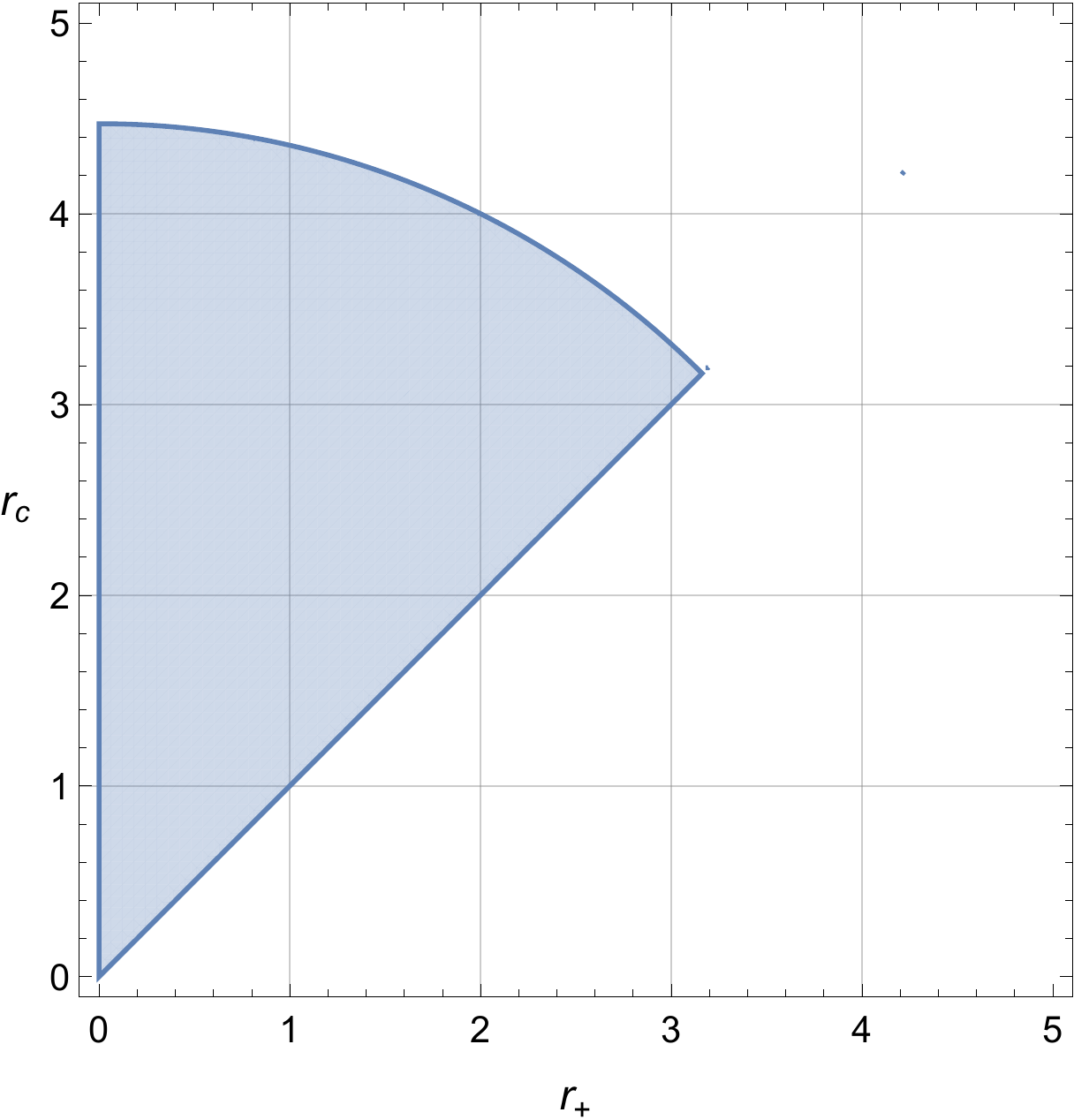}
	\caption{Thermodynamic volume of the uncharged Gauss-Bonnet black hole in $D=5$ with $\lambda_{GB}=0.1$ and $\Lambda=0.3$. The blue shaded region indicates positivity of the volume $V$. The diagonal and curved boundaries mark, respectively, where $r_c=r_+$ and $r_c=r_{cosmo}$.}
	\label{Fig1}
\end{figure}

In Figure~\ref{Fig1} we plot regions where  the thermodynamic volume is positive, as a function of $r_+$ and $r_c$ with $\Lambda$ and $\lambda_{GB}$ held fixed. The volume is positive in the blue shaded region. There are two boundaries enclosing this region. The diagonal line marks the boundary where $r_c>r_+$; since we are restricting the cavity to lie outside the event horizon, we are automatically on the upper-left side of this line. The second, curved boundary represents the line along which $r_c=r_{cosmo}$. Again we are restricting our cavity to lie within the cosmological horizon, therefore the thermodynamic volume is positive in all regions of interest. This picture is qualitatively identical in higher dimensions.

\subsection{5 dimensional black holes}

Having derived the general results for the uncharged Gauss-Bonnet black holes, we turn to the study of their phase structure, starting with $D=5$. The quantity of interest is the {\it free energy} $F=E-TS$, the quantity that is minimized by the equilibrium state of the system.

\begin{figure}[h]
	\includegraphics[width=0.49\textwidth]{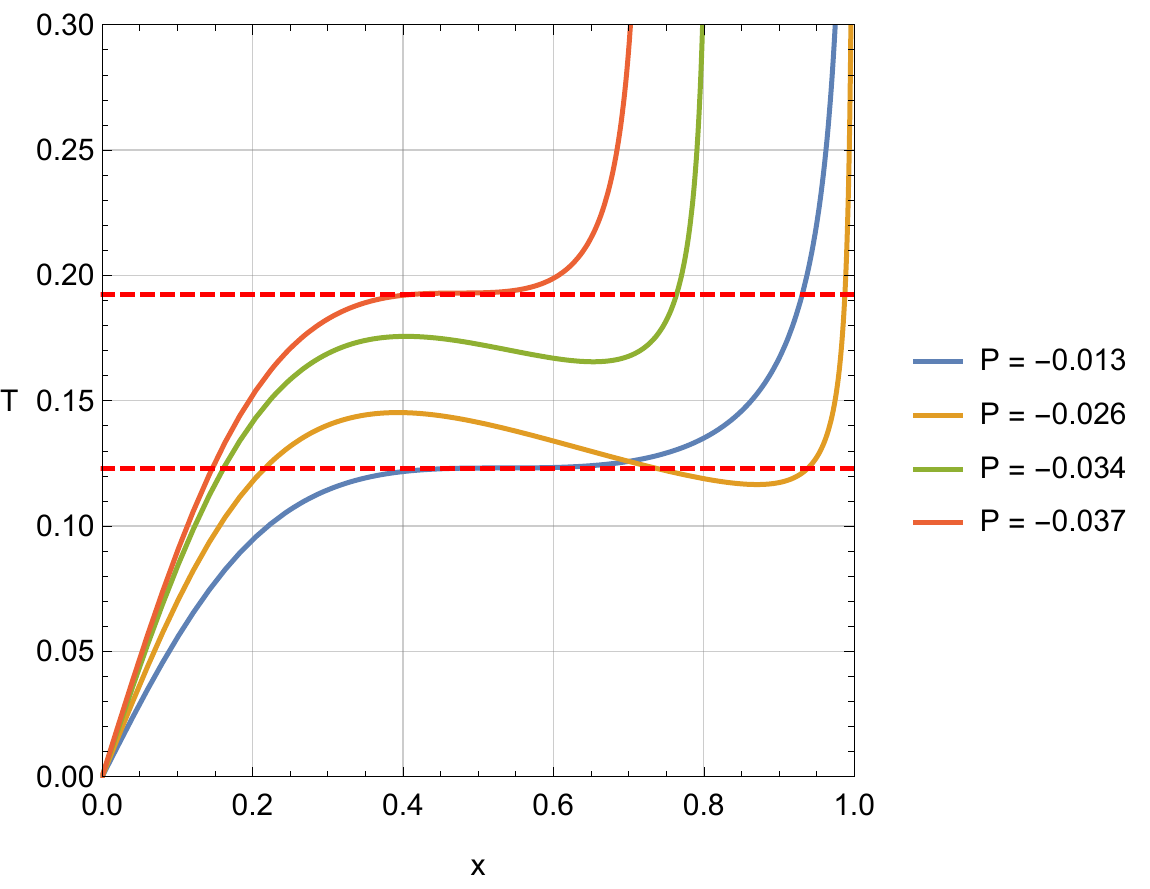}\quad\includegraphics[width=0.49\textwidth]{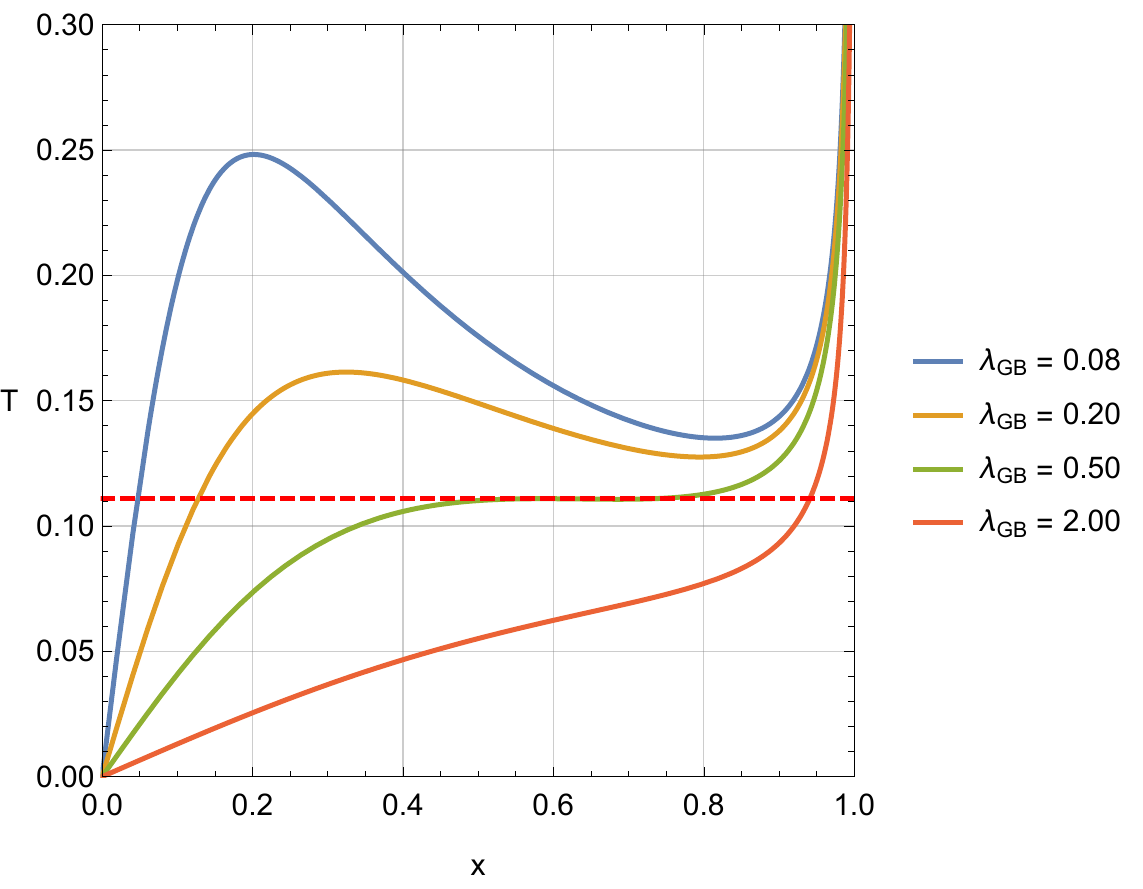}
	\caption{Temperature $T$ as a function of $x=r_+/r_c$ for fixed cavity radius $r_c=2$ and $D=5$, showing regions where $r_+$ is multi-valued, signaling a possible phase transition. \textbf{Left:} Varying pressure with $\lambda_{GB}=0.3$. 
	The red dashed horizontal lines demarcate the region in which $T$ is not a monotonically increasing
	function of $x$. \textbf{Right:} Varying Gauss-Bonnet coupling with $P=-0.02$. Above the red dashed line $T$ is not a monotonically increasing
	function of $x$.	}
\label{Fig2}	
\end{figure}

One way to examine the behaviour of this system is to realize that for a phase transition to occur at a given temperature $T$, the function $r_+(T)$ with fixed $\{r_c,\lambda_{GB},P\}$ must be multi-valued.\footnote{This is true throughout most of the parameter space, however there are some points where the temperature may be single-valued but at an inflection point, signaling a second-order phase transition.} The interpretation of this is the existence of multiple thermodynamically competing states with equal temperature but different horizon radii $r_+$, which will in general have different free energies. The transition of the horizon radius being a single valued function of the temperature to multi-valued thus (typically) corresponds to the free energy becoming multi-valued at fixed temperature. This will not, however, tell us about the stability of the phases or nature of the transition, which must be determined from the free energy itself. An analytic study of the roots of $T(r_+)$ is not possible in this case since the expression does not admit a closed form solution for $r_+$. In Figure~\ref{Fig2} we plot the temperature as a function of $x\equiv r_+/r_c$ for fixed coupling $\lambda_{GB}$ and varying pressure $P$, as well as fixed $P$ and varying $\lambda_{GB}$, showing the transition from the single to multi-valued regime.

On the left in Figure~\ref{Fig2}, one can see that at fixed coupling there is a compact region $[P_{min},P_{max}]$
between the red dashed lines where $T$ is not a monotonically increasing function of $x$,  in which case the horizon radius is a multi-valued function of the temperature. Below the minimum and above the maximum pressure, there is only one thermodynamically allowed state. In contrast, on the right we see that at fixed pressure, there is a maximum value of the Gauss-Bonnet coupling below which the horizon radius is multi-valued. However there is no minimum and the phase transition (if it exists for a given choice of parameters at fixed pressure) always persists in the limit $\lambda_{GB}\rightarrow 0$. This type of analysis gives hints as to which regions in parameter space may have multiple competing phases, but only the free energy $F$ can tell the whole story. The free energy of empty de Sitter space is $F=0$, so even if multiple black hole phases exist, they may not be thermodynamically preferred, as we will see.
\\
\begin{figure}[h]
	\includegraphics[width=0.49\textwidth]{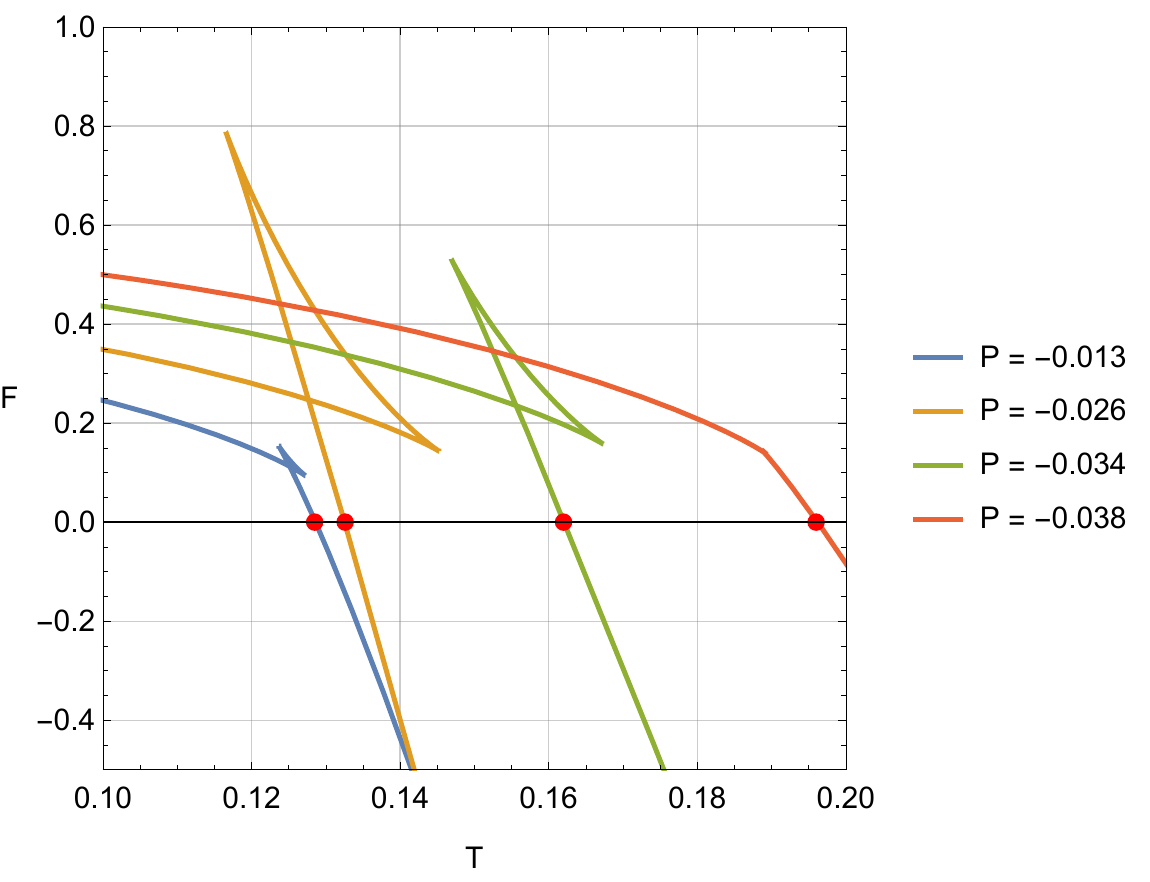}\quad\includegraphics[width=0.48\textwidth]{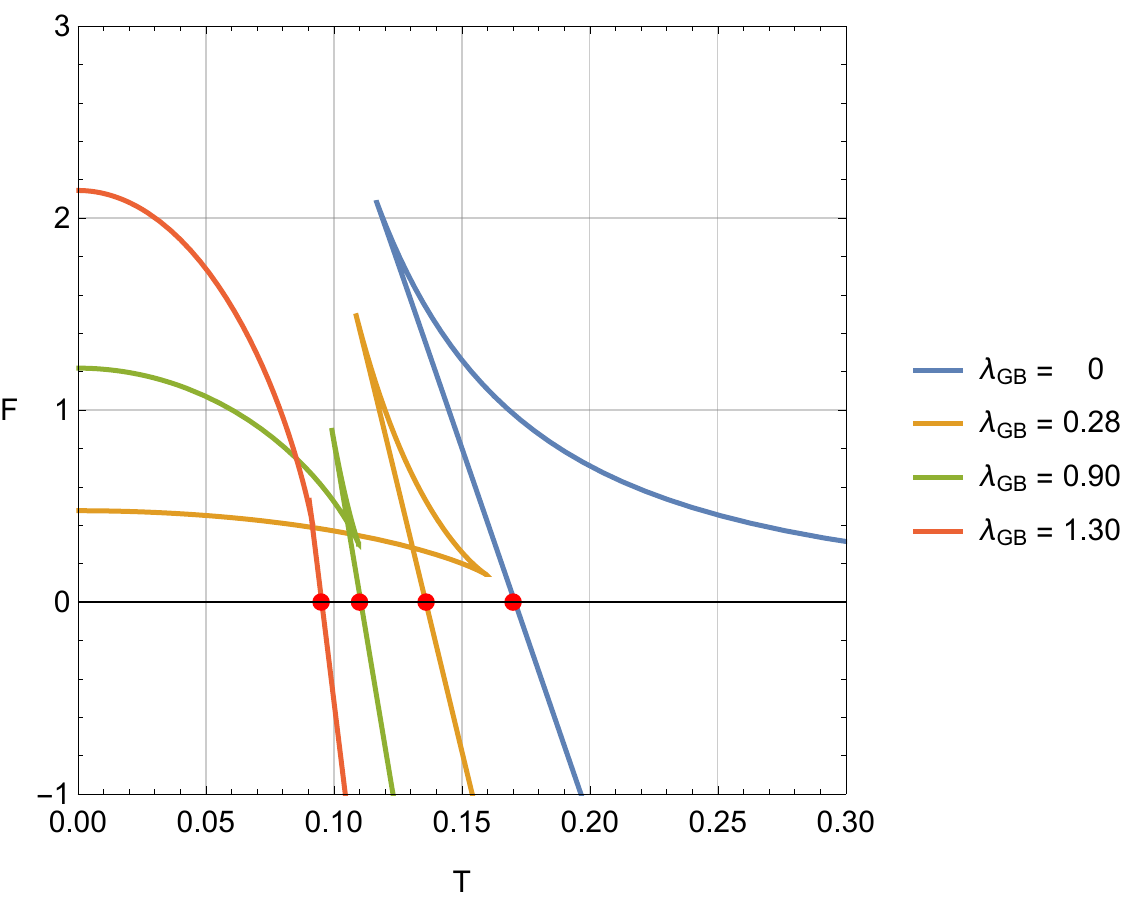}
	\caption{Free energy of the Gauss-Bonnet-de Sitter black hole in $D=5$ with $r_c=2$, showing a Hawking-Page phase transition from radiation to a large black hole, where the free energy crosses $F=0$. \textbf{Left:} Varying pressure with $\lambda_{GB}=0.3$. \textbf{Right:} Varying Gauss-Bonnet coupling with $P=-0.03$. For very small values of $\lambda_{GB}$ the free energy limits to the Einstein case where there is a Hawking-Page phase transition with a minimum black hole temperature.	}
	\label{Fig3}	
\end{figure}

In Figure~\ref{Fig3} we plot the free energy $F=E-TS$ parametrically as a function of $T$ with $r_+$ as the parameter. The thermodynamically preferred state is the one that globally minimizes the free energy. As the temperature of the system increases, the system will follow the line with lowest free energy whenever a crossing is reached. On the left of Figure~\ref{Fig3}, we plot the free energy at fixed value of the Gauss-Bonnet coupling and varying pressure. There is a crossing of the black hole free energy with itself, corresponding to a first order small-to-large black hole phase transition. However, since this crossing is above the free energy of thermal de Sitter space ($F=0$), the relevant transition occurs at the red dots where the black hole free energy line crosses $F=0$. This represents a Hawking-Page phase transition from radiation, or thermal de Sitter space, to a large black hole, and is generically seen in asymptotically AdS black holes.

We briefly clarify the notion of the Hawking-Page transition in de Sitter space. Here the transition is between a thermal gas confined to the cavity (this is what we mean by `thermal de Sitter')  and a black hole confined to the cavity. The temperature of both the gas and the black hole will generically be different from the temperature associated with the cosmological horizon. This is justified by the presence of the cavity: the boundary conditions imposed by the cavity allow for the control of the temperature of the cavity and its contents independently of the cosmological horizon.

On the right side of Figure~\ref{Fig3}, we plot the free energy at fixed pressure for varying coupling $\lambda_{GB}$. Here we see that below a critical value of the coupling (in this case $\lambda_{GB}\sim1.3$), a crossing forms in the black hole free energy, though because it is always above $F=0$, we again only have a Hawking-Page phase transition where the large black hole branch crosses $F=0$. In the limit $\lambda_{GB}\rightarrow 0$, corresponding to Einstein gravity, we recover the results of ~\cite{Simovic:2018tdy}. Note that while the Hawking-Page transition is present for any choice of $\lambda_{GB}$, three situations are distinguished by the number of unstable phases available to the system. When $\lambda_{GB}=0$, there are two black hole phases, when $0<\lambda_{GB}<1.3$ there are three black hole phases, and when $\lambda_{GB}>1.3$ there is only one. The presence of the Gauss-Bonnet correction also gives rise to unstable black hole phases down to $T=0$, while in the Einstein limit $\lambda_{GB}\to 0$ there is a minimum temperature black hole where the free energy reaches a point.

\subsection{6+ dimensional black holes}

\begin{figure}[h]
	\includegraphics[width=0.49\textwidth]{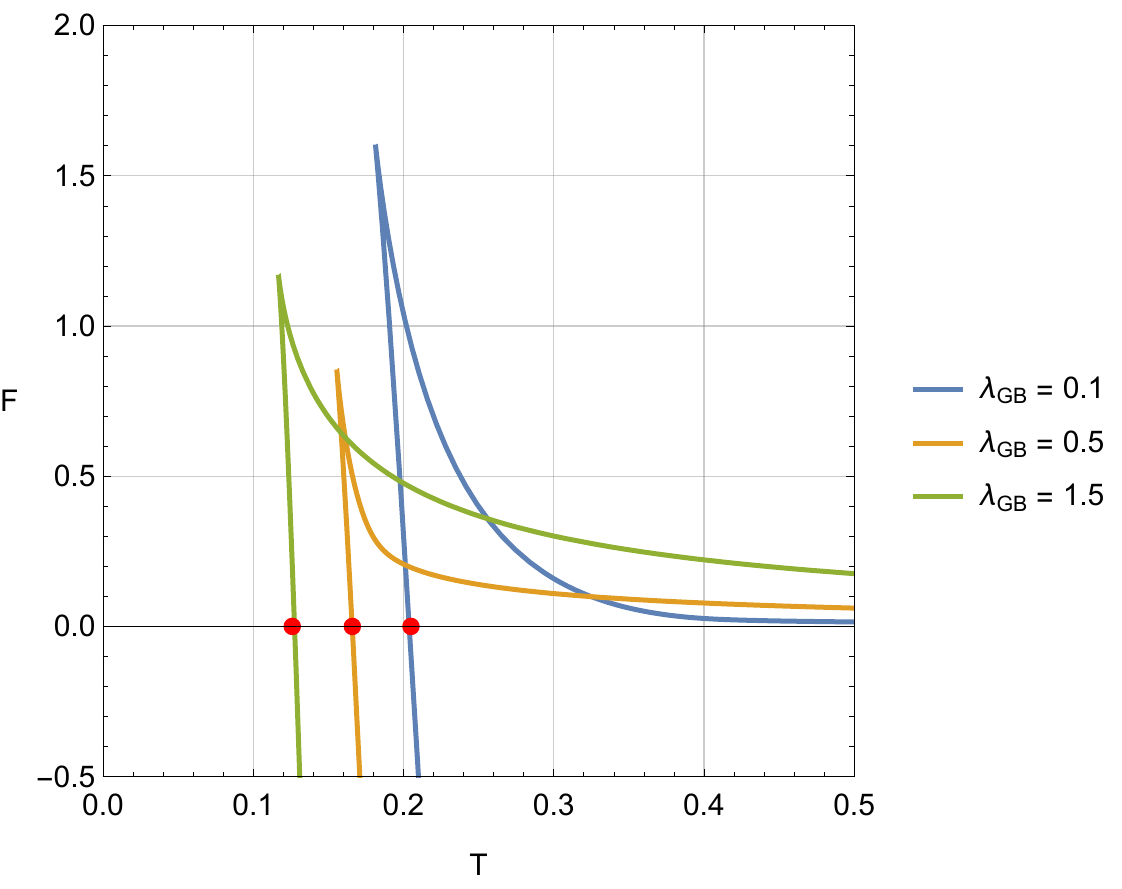}\quad\includegraphics[width=0.48\textwidth]{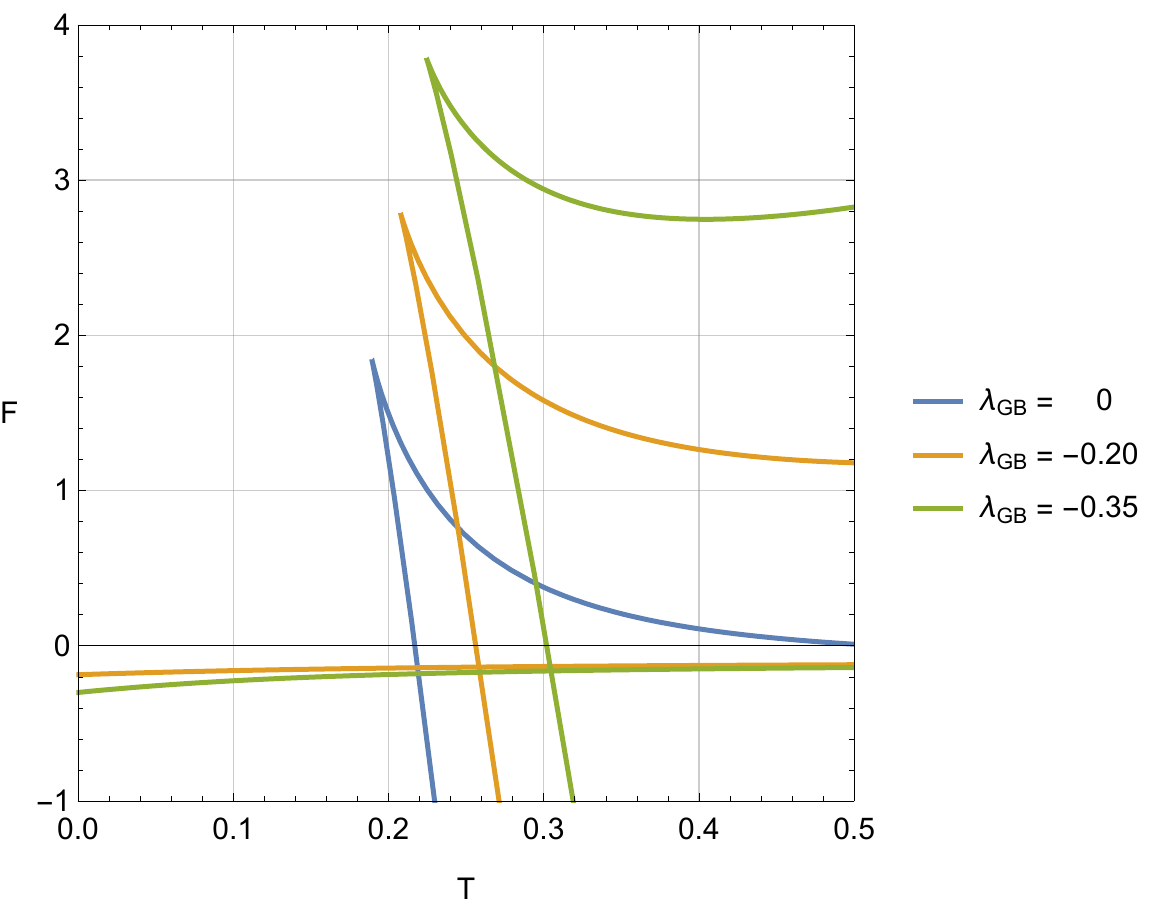}
	\caption{Free energy of the Gauss-Bonnet-de Sitter black hole with $D=6$, $r_c=2$, $P=-0.03$, and varying Gauss-Bonnet coupling. \textbf{Left:} With $\lambda_{GB}>0$, there is a first-order phase transition from thermal de Sitter to a large black hole. \textbf{Right:} With $\lambda_{GB}<0$, there is a first-order small-large black hole phase transition. Note the free energy of the small black hole branch is below that of radiation in this case. This behaviour is qualitatively the same in higher dimensions.}
	\label{Fig4}
\end{figure}

When $D>5$, there are two cases of interest. When $\lambda_{GB}>0$, there is again only a Hawking-Page transition from thermal de Sitter space to a large black hole, with a minimum black hole temperature as in the Einstein limit of the $D=5$ case. This is encoded in the fact that the temperature is never more than double-valued as a function of $x=r_+/r_c$. When $\lambda_{GB}<0$ however, we observe a small-large black hole phase transition, since the small black hole branch now has lower free energy than the radiation phase, and is the preferred state of the system at low temperatures. We demonstrate this in Figure~\ref{Fig4}, plotting the free energy for fixed pressure and varying coupling for $D=6$. Note that unlike in the $D=5$ cases, when the coupling is negative, the small black hole branch (represented by the near-horizontal lines)  has free energy less than that of radiation, and is not continuously connected to the large black hole branch. The phase structure is qualitatively identical for higher dimensions, with only the precise value of the critical temperature differing for a given choice of cavity size, pressure, and coupling.

As is well-known, at sufficiently negative coupling Gauss-Bonnet gravity exhibits pathological behaviour such as naked singularities or negative string tension (if viewed as arising from $\alpha'$ corrections in string theory)~\cite{Boulware:1985wk}. However, small negative couplings cannot be completely ruled out via analysis of physicality conditions~\cite{Buchel:2009sk}. Note also that when one considers the dynamical stability of these black holes, the coupling is further constrained~\cite{Cuyubamba:2016}, and that these black holes will be dynamically unstable in higher dimensions~\cite{Konoplya:2009}. Here, we remark only on the fact that a change in sign of the coupling leads to very different phase structure. 

\section{Charged Gauss-Bonnet Black Holes}

We next consider the inclusion of a $U(1)$ gauge field in the action \eqref{action!}, and study the resulting thermodynamics. In general, the presence of a Maxwell-like field leads to re-entrant phase transitions and van der Waals-like behaviour. This is tied to the $r^{6-2D}$ falloff of the charge term that appears in the metric function when such a field is present, leading to additional roots in the temperature. The action is the same as before,
\begin{align} 
I_* &= \frac{(D-2)\Omega_{D-2} \beta_c  r_c^{D-5}}{24 \pi G} \left[\sqrt{f(r_c)} \left( -3 r_c^2 + 2 \lambda_{\rm GB}(f(r_c) - 3) \right) \right.
\nonumber\\
&\left.- \sqrt{f_0(r_c)}\left( -3 r_c^2 + 2 \lambda_{\rm GB}(f_0(r_c) - 3) \right) \right] - \frac{\Omega_{D-2} r_+^{D-2}}{4 G} \left[1 + \frac{2(D-2)\lambda_{\rm GB}}{(D-4) r_+^2} \right]  \, ,
\end{align}
however now the metric function takes the form
\be
f(r) = 1+ \frac{r^2}{2 \lambda_{GB}} - \frac{r^{2 - D/2} \sqrt{r^D+4\,\lambda_{GB} \big(r\, \omega_{D-3} -q^2\, r^{4-D}-r^D/L^2\big)}}{2 \lambda_{GB}}  \,.
\ee
The energy $E$, temperature $T$, and entropy $S$ therefore take the same functional form as in the uncharged case, namely \eqref{f0}--\eqref{temp}, differing only by the form of the metric function $f(r)$. The free energy $F=E-TS$ thus also takes the same general form as in the uncharged case.

\subsection{The first law}

When charge is present, the first law must be supplemented by an additional $\phi\, dQ$ term, with $\phi$ representing the electric potential of the spacetime measured at the cavity, and $Q$ the total charge:
\be
dE=TdS+VdP+\sigma dA+\phi\, dQ+\Phi_{GB}\lambda_{GB}
\ee
We can again derive the conjugate quantities $\{V,\phi,\sigma,\lambda_{GB}\}$ by enforcing the first law. We omit the expressions here since they are long and offer no particular insight. They are structurally similar to those in Section 3.1, with the addition of a number of $q$-dependent terms. One can also verify that with these quantities the Smarr relation holds, which in the presence of charge reads:
\be
(D-3)E=(D-2)TS+(D-3)\phi\, Q-2PV+2\Phi_{GB}\lambda_{GB}\, .
\ee

\begin{figure}[h]
	\centering
	\includegraphics[width=0.45\textwidth]{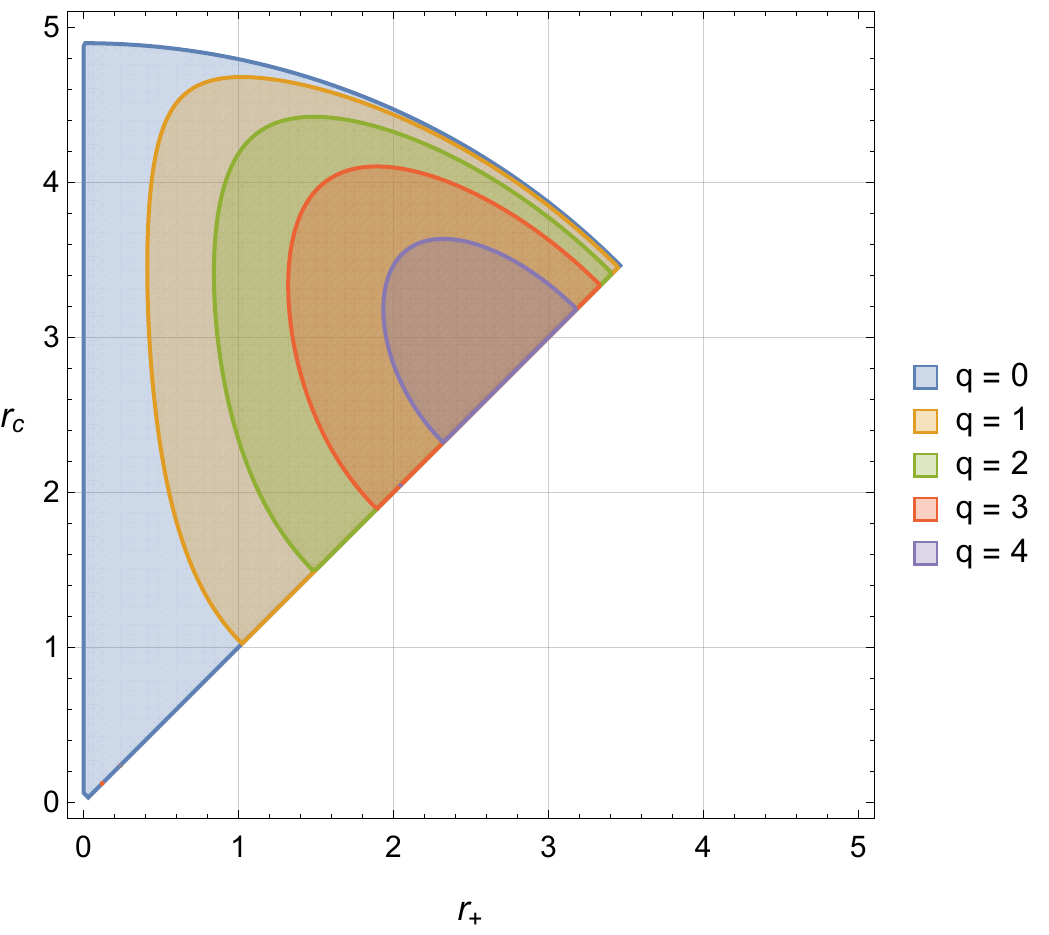}
	\caption{Thermodynamic volume of the charged Gauss-Bonnet black hole in $D=5$ with $\lambda_{GB}=0.1$, $\Lambda=0.25$, and varying charge. The shaded regions indicate positivity of the volume $V$. The diagonal boundary marks where $r_c=r_+$. }
	\label{Fig5}
\end{figure}

As before, we examine the thermodynamic volume to check for positivity. In Figure~\ref{Fig5}, we plot regions where $V>0$ for varying charge $q$.

When $q=0$ we reproduce Figure~\ref{Fig1}. With non-zero $q$, regions where the thermodynamic volume is positive are smaller than in the uncharged case. The inner boundary of the shaded regions correspond to the location of the inner horizon $r_-$ of the charged black hole. Outside of these regions, the volume is not negative, but rather imaginary. Since we are restricting the cavity to lie within $r_-<r_c<r_{cosmo}$, we have an everywhere positive thermodynamic volume as in the uncharged case.

\subsection{Phase structure}

\begin{figure}[h]
	\includegraphics[width=0.49\textwidth]{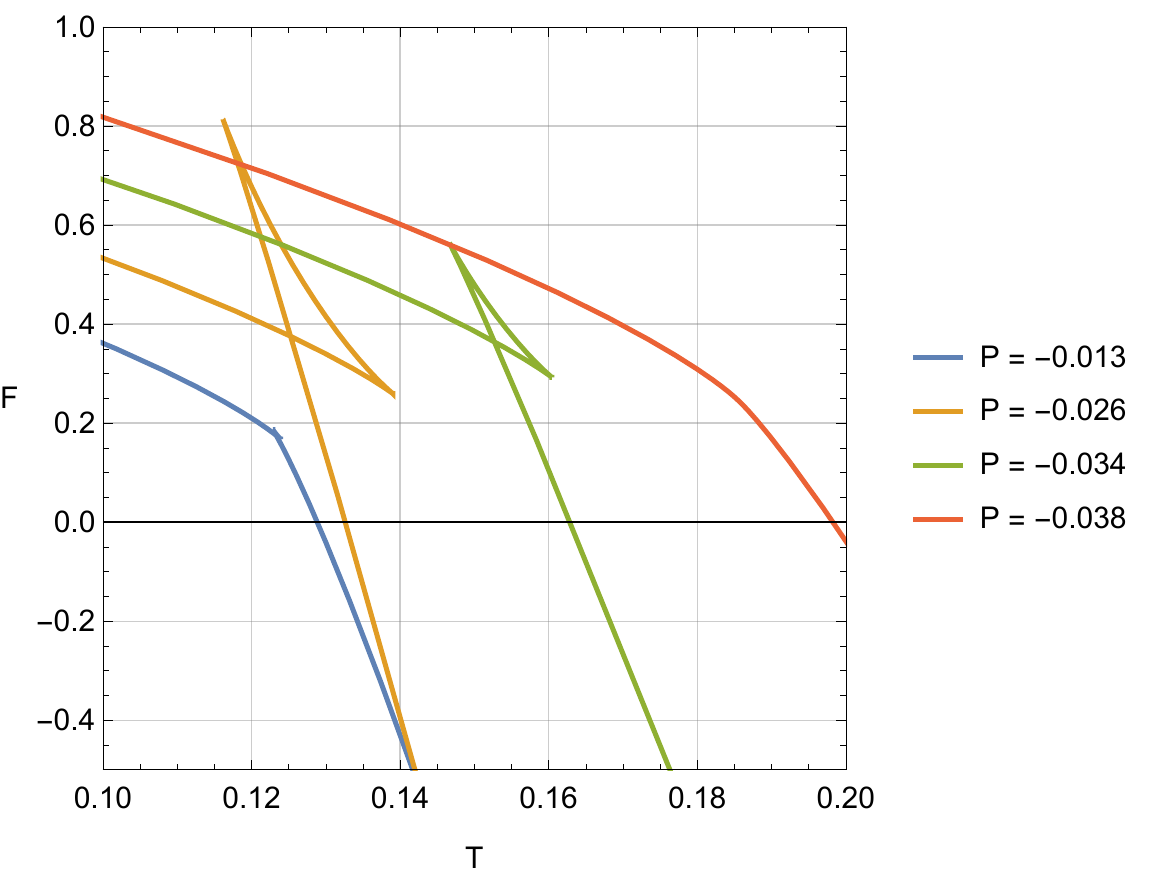}\quad\includegraphics[width=0.49\textwidth]{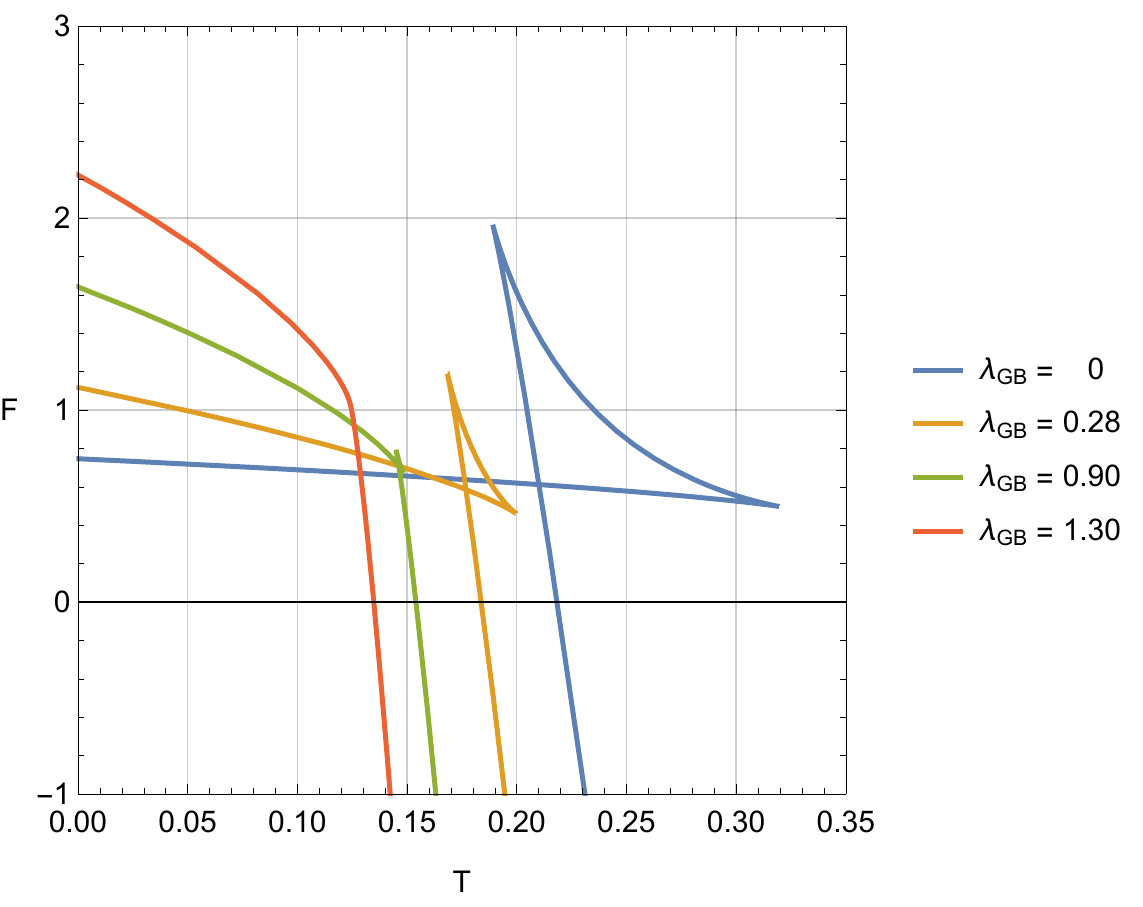}
	\caption{Free energy of the charged Gauss-Bonnet-de Sitter black hole with $D=5$ and $r_c=2$, showing a first-order phase transition from a small black hole to a large black hole. \textbf{Left:} Varying pressure with $\lambda_{GB}=0.3$. \textbf{Right:} Varying Gauss-Bonnet coupling with $P=-0.03$. Note that in the $\lambda\to 0$ limit we have a small-large phase transition, as opposed to the Hawking-Page transition that emerges in this limit in the uncharged case.}
	\label{Fig6}
\end{figure}

We again turn to an analysis of the free energy $F=E-TS$ to uncover the phase structure, this time for charged Gauss-Bonnet-de Sitter black holes. Figure 6 shows a plot of the free energy of the black hole both for varying pressure and varying coupling. While the free energy of Figure 6 looks identical to the uncharged case, the interpretation is different in an important way. Since we are working in the canonical ensemble, the charge $q$ of the black hole is fixed. This means that there is no Hawking-Page phase transition at the crossing of the black hole free energy with the $F=0$ line, because a black hole cannot evaporate while its charge is held fixed. Instead, we have a small-large black hole phase transition where the black hole free energy line crosses itself. The system will follow the branch with lowest free energy, so the black hole suffers a jump from small $r_+$ to large $r_+$ at the crossing. With the cavity present, there is not only a minimum critical pressure $P_{min}$ below which a kink forms in the free energy, but also a maximum pressure $P_{max}$. These critical pressures coincide with the values of $[P_{min},P_{max}]$ at which $x$ becomes multi-valued at a given temperature.  At these critical pressures, a cusp forms in the free energy where a second order phase transition occurs, as in the red lines of Figure~\ref{Fig6}. Outside of the range $[P_{min},P_{max}]$, the free energy is monotonic and there is no phase transition. 
\\

On the right side of Figure~\ref{Fig6}, we vary instead the Gauss-Bonnet coupling with pressure held fixed. For couplings above a certain value (in this case $\lambda_{GB}=1.3$) there is only one phase. Below this value, a crossing forms and we have a small-large phase transition. Notably, in the Einstein limit $\lambda_{GB}\to0$ this small-large transition persists, unlike in the uncharged case considered previously. Unlike the uncharged case, when charge is present all of the qualitative features of the $D=5$ black hole remain the same in higher dimensions; only the precise values of the free energy and critical points change, but all other phase structure and limiting behaviour is identical.

\begin{figure}[h]
	\centering
	\includegraphics[width=0.75\textwidth]{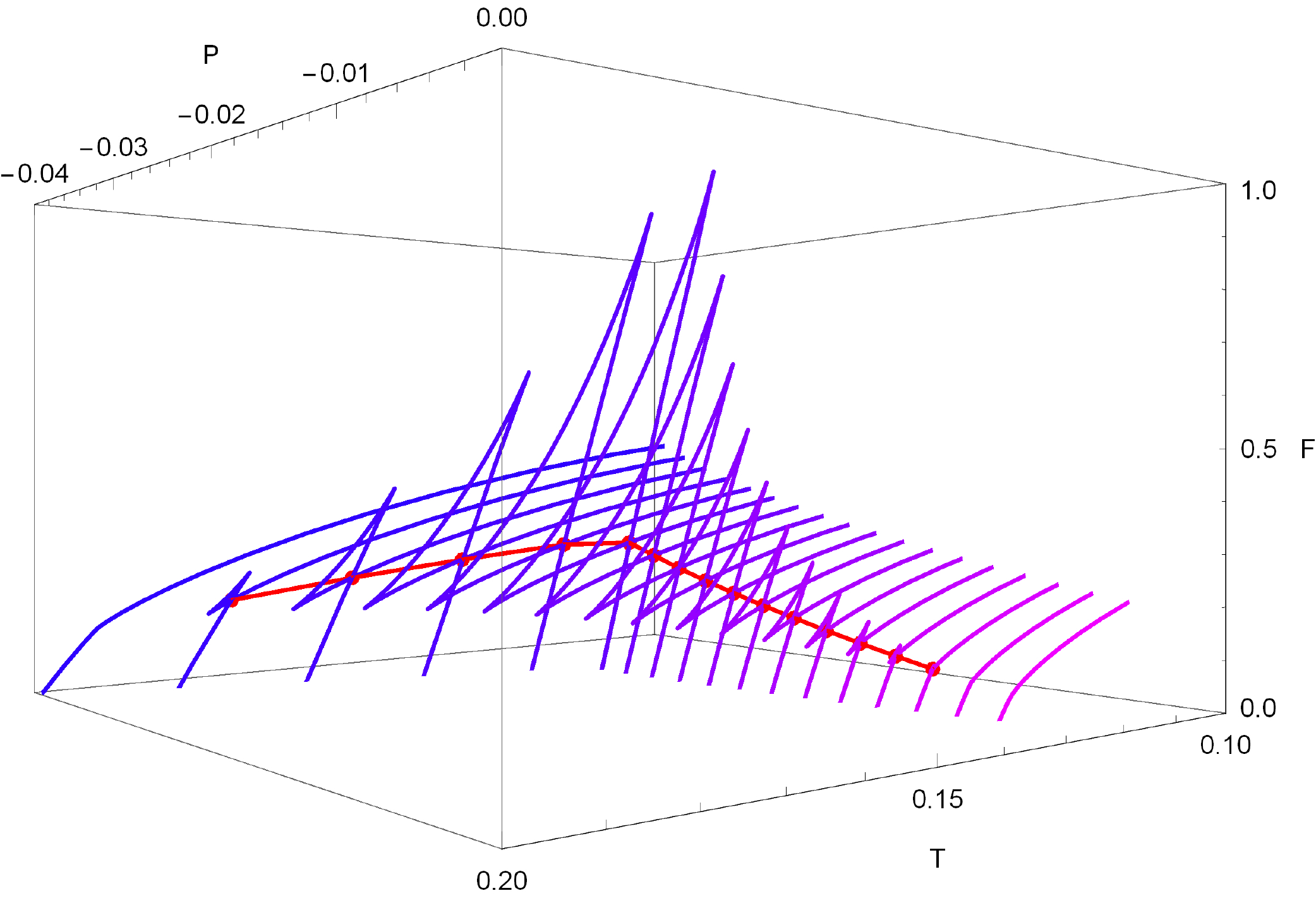}
	\caption{Free energy of the Gauss-Bonnet-de Sitter black hole in $D=5$ with $r_c=2$ and $\lambda_{GB}=0.3$, showing the formation of a swallowtube corresponding to a compact region of first-order phase transitions from a small to large black hole. Each line corresponds to a constant-pressure slice, while red dots mark the location of the critical temperature within each slice. The red line is the coexistence line.}
	\label{Fig7}
\end{figure}

\begin{figure}[h]
	\centering
	\includegraphics[width=0.51\textwidth]{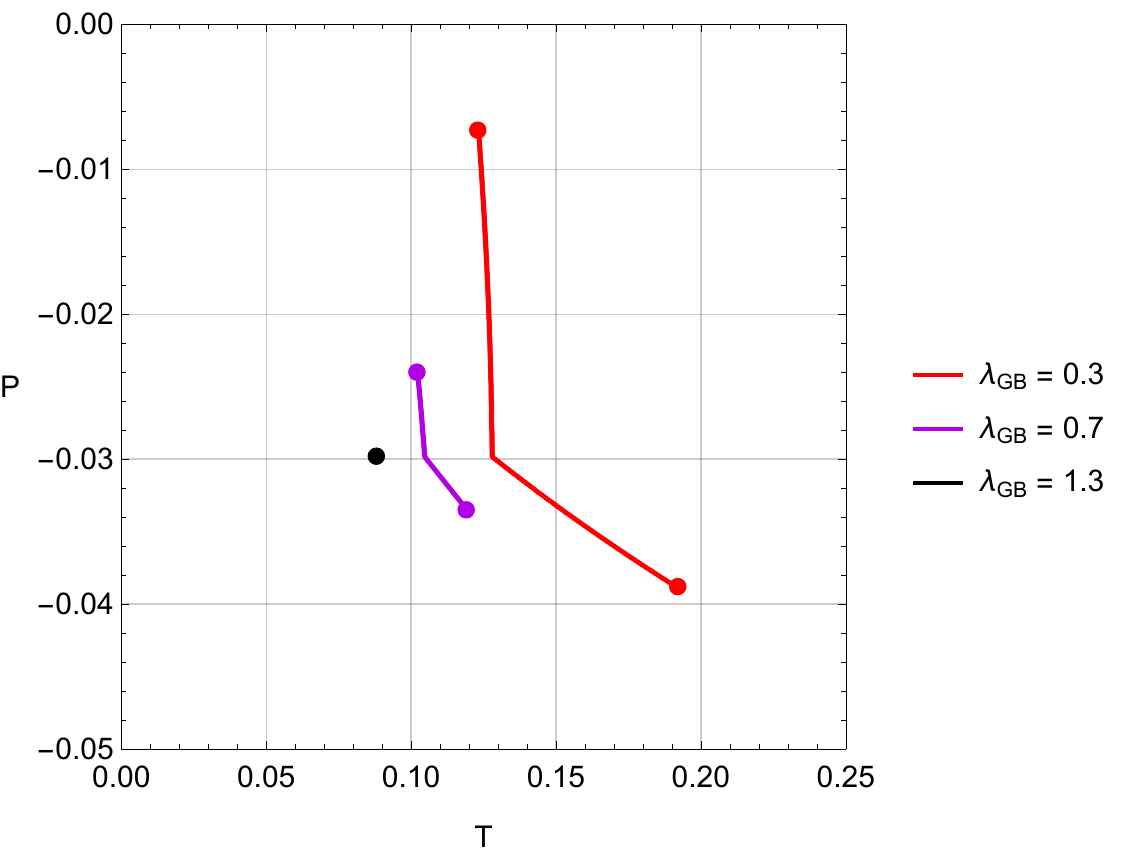}\qquad\includegraphics[width=0.405\textwidth]{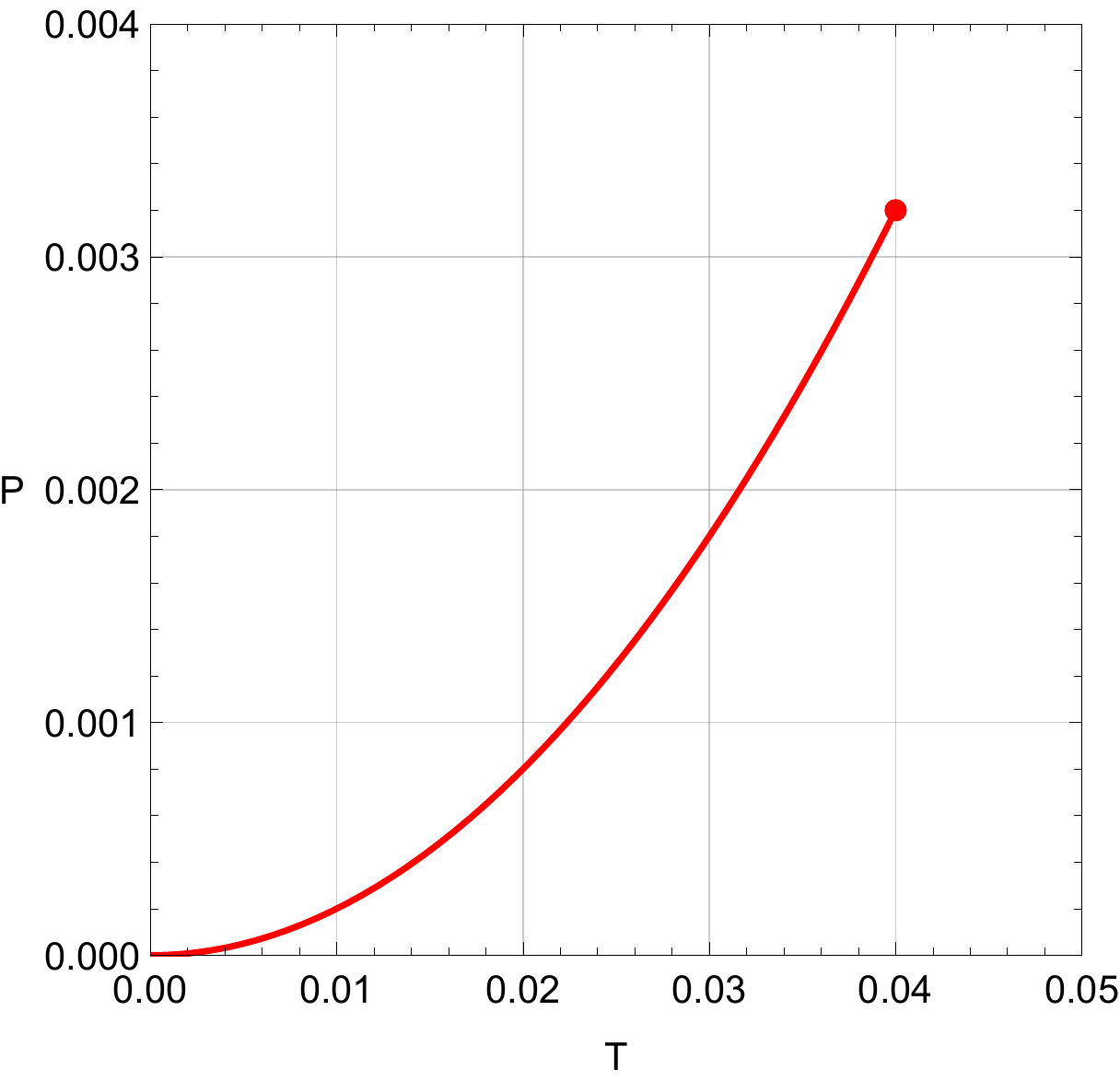}
	\caption{Coexistence curves for black holes, along  which the small and large black hole phases coexist. \textbf{Left:} The uncharged Gauss-Bonnet black hole with $r_c=2$ and varying $\lambda_{GB}$. The large dots mark the critical pressures $P_{min}$ ($P_{max}$), above (below) which there is no phase transition. At these points a second order phase transition from small to large black hole transition occurs.	 \textbf{Right:} The uncharged AdS black hole, illustrating the difference when there is no cavity present. There is only one second order phase transition at the red dot $P_{max}$.}
	\label{Fig8}
\end{figure}

\begin{figure}[h]
	\centering
	\includegraphics[width=0.75\textwidth]{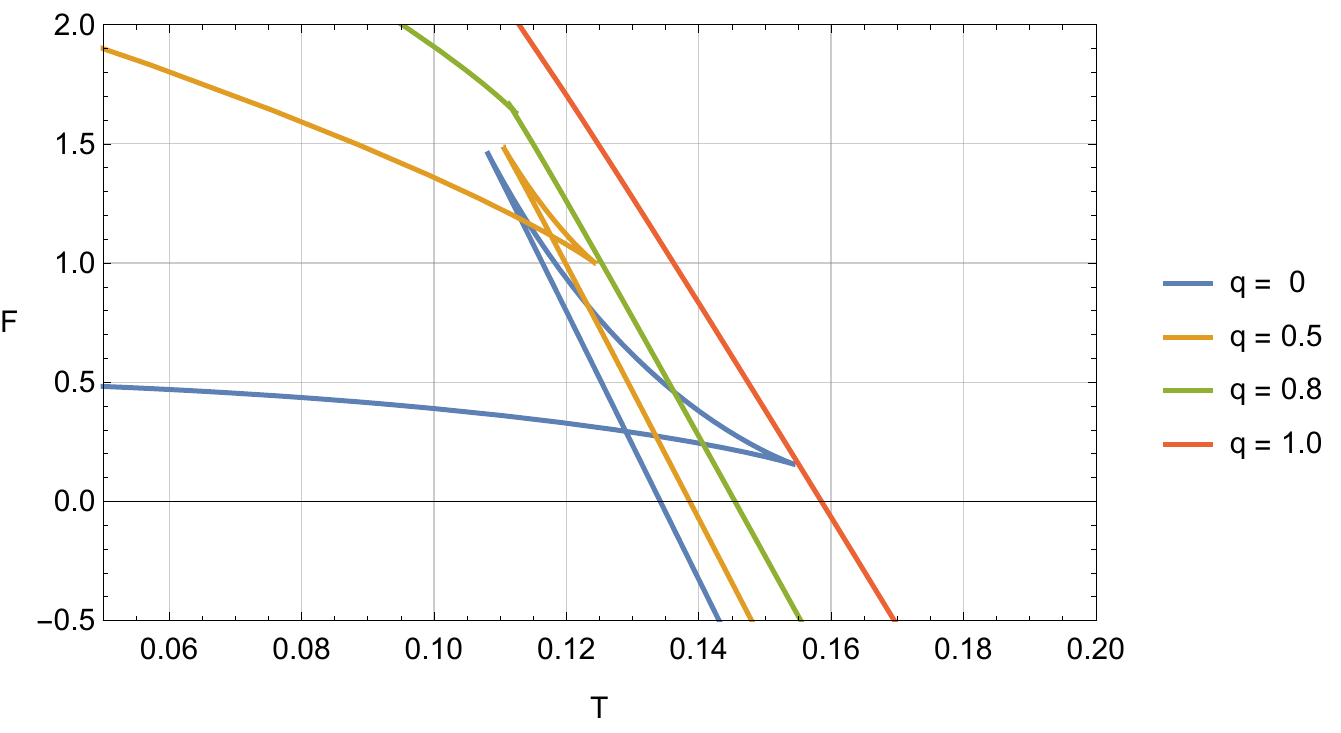}
	\caption{Free energy of the charged Gauss-Bonnet-de Sitter black hole with $D=5$, $r_c=2$, and $\Lambda=0.3$, showing a first-order phase transition from a small black hole to a large black hole below a critical value of the charge $q$. In this case, $q_c\sim0.8$.}
	\label{Fig9}
\end{figure}

Unlike the typical `swallowtail' behaviour seen in asymptotically AdS black holes (see for example~\cite{KubiznakMann:2012}), the free energy here forms a tube in $F-T-P$ space, as shown in Figure~\ref{Fig7}. This `swallowtube' behaviour, first observed in \cite{Simovic:2018tdy}, is in stark contrast to the swallowtails that arise in black hole systems without cavities. In those systems, there is only a maximum pressure $P_{max}$ below which the phase transition is present. Here, there is also a minimum pressure  $|P_{min}|>0$  that is reached where another second-order phase transition occurs, and only between these two pressures is there a small-large black hole phase transition. In Figure~\ref{Fig8}, we plot the coexistence curve for the black hole and compare it to a typical AdS black hole. These curves are lines in $P-T$ space along which the small and large black hole phases simultaneously exist and have equal free energy. Note the striking difference: when a cavity is present, the coexistence line terminates at two second-order phase transitions as opposed to one. One can also vary the charge $q$ at fixed values of the pressure and coupling, as shown in Figure~\ref{Fig9}. Here we see a single critical value of the charge $q_c$ for a given choice of $\Lambda$ and $\lambda_{GB}$, below which there exists a small-large black hole phase transition, persisting down to $q=0$. Notice that swallowtubes only exist in $F-T-P$ space: both in $F-T-q$ and $F-T-\lambda_{GB}$ space we see a swallowtail instead, with just one critical value of the respective parameters $q$ or $\lambda_{GB}$.

\section{Conclusions}

We have studied the phase structure of both charged and uncharged de Sitter black holes in Gauss-Bonnet gravity, in the canonical ensemble. The presence of an isothermal cavity, equivalent to fixing boundary value data on a finite surface in the spacetime, allows us to have a notion of thermodynamic equilibrium in these asymptotically de Sitter spacetimes, which normally are not in equilibrium due to the two horizons present. What we have seen is a host of interesting phenomena. In the uncharged case, there exist Hawking-Page-like phase transitions throughout most of the parameter space, with a number of unstable black hole phases present. In the special case of $\lambda_{GB}<0$, we find a region of first order small-large black hole phase transitions, where the free energy of the small black hole branch becomes smaller than that of radiation. Interestingly, while exotic re-entrant phase transitions and triple points are seen in 6-dimensional uncharged Gauss-Bonnet black holes in AdS spacetimes, here we see only a Hawking-Page phase transition in the 6-dimensional case. This touches on an important point: that while anti-de Sitter space acts like a `box' that confines radiation much like a cavity does (allowing the black hole to reach thermodynamic equilibrium), these two methods of achieving equilibrium leave their imprint on the phase structure. One cannot understand the thermodynamic behaviour of a black hole without also considering how it is being maintained at equilibrium, for the exact method by which this is achieved affects significantly the resulting behaviour, even if the mechanisms seem qualitatively alike.

When charge is present, the story is considerably different. We generically see first order small-large black hole phase transitions encoded in the presence of a swallowtube in the $F-T-P$ space, with second order phase transitions at the minimum and maximum pressures representing the end points of the tube. This swallowtube behaviour appears to be a characteristic feature of black holes embedded in isothermal cavities~\cite{Simovic:2018tdy}. Interestingly, such tubes only exist in $F-T-P$ space. When either the charge $q$ or coupling $\lambda_{GB}$ are varied, only a swallowtail emerges. These parameters do however control the size of the swallowtube in $F-T-P$ space, and for any particular choice of $P$ for which a tube exists, one can find values of $q$ and $\lambda_{GB}$ such that the two ends of the tube (corresponding to a second order phase transition) meet. Based on previous works~\cite{Dolan:2014vba} we would expect that the merging of two critical exponents would yield novel critical exponents. However, the investigation of this expectation is difficult here as there is no first-order phase transition present in the case where the `merged' critical exponent occurs --- it is a truly isolated second-order phase transition. It is an open question as to whether different critical exponents emerge at this point, and how universal such behaviour is when a cavity is present.

\section{Acknowledgements} This work was supported in part by the Natural Sciences and Engineering Research Council of Canada.

\section*{Appendix A}

Here we present expressions for the conjugate variables derived from the first law for both uncharged and charged Gauss-Bonnet-de Sitter black holes. The first law takes the general form:
\be
dE=TdS+VdP+\sigma dA+\phi\,dQ+\Phi_{GB}\lambda_{GB}\,.
\ee
For convenience, we define
\be
\gamma=\dfrac{-2\Lambda}{(D-1)(D-2)}
\ee
and let $f(r)\rightarrow f$ and $f_0(r)\rightarrow f_0$, though note that these functions depend on $(r_+,r_c,\gamma,\lambda_{GB},q)$. Using (3.7)--(3.8) in the first law we can determine the following:
\begin{align}
V=\ &\frac { 2\, \Omega_{D-3} \left[\Big( \lambda_{GB} \left( f_0 -1 \right) r_c^{D}-\tfrac{1}{2}\,r_c^{D+2} \Big) \frac{\partial f_0}{\partial \gamma} \sqrt {f\,}-\Big( r_c^{D}\lambda_{GB}(f-1)-\tfrac{1}{2}\,r_c^{D+2} \Big)\frac{\partial f}{\partial \gamma}  \sqrt {f_0 }\, \right] }{	r_c^5\sqrt {f_0\, f\, }\left( D-1 \right) }\\
\sigma=\ &\frac {(D-2)\,\Omega_{D-3}}{96\pi^{2} \sqrt {f_0\,f\,}\,r_c^7}\,\bigg[r_c^{D}\sqrt {f_0}\,\bigg(f \left( \lambda_{GB} \left( d-5 \right) f + \tfrac{3}{2}\left(3-D\right) r_c^{2}-3\lambda_{GB} \left(D-5\right) \right)\nonumber\\
&+\tfrac{3}{2}\, \left( {\lambda}_{GB} f -\tfrac{1}{2}r_c^{2}-\lambda_{GB} \right) r_c\,{\tfrac {\partial f}{\partial r_c}} \bigg)+\bigg(r_c^{D+2}\left(\tfrac{3}{4}\left( {\tfrac {\partial f_0}{\partial r_c}}  \right) r_c+\tfrac{3}{2}\,f_0 \left( D-3 \right)  \right) \bigg) \sqrt {f\,}\\
& -r_c^{D} \left( \tfrac{3}{2}\,r_c\, \left( f_0 -1 \right) {\tfrac {\partial f_0}{\partial r_c}} +f_0 \left(D-5 \right)  \left( f_0 -3 \right)  \right) \lambda_{GB}\bigg]\nonumber\\
\nonumber\\
\Phi_{GB}=\ &\frac { (D-2)\,\Omega_{D-3}}{12\pi\sqrt {f_0\,f\, }r_c^{5}}  \, \bigg[ \bigg(\tfrac{3}{2}\Big(\,r_c^{D}\lambda_{GB}(1-f_0)+\tfrac{1}{2}\,r_c^{D+2} \Big) {\tfrac {\partial f_0}{\partial \lambda_{GB}}} -r_c^{D}f_0  \left(f_0 -3 \right)  \bigg) \sqrt {f }\nonumber\\
&+\left(  \tfrac{3}{2}\left( f\,\lambda_{GB} -\tfrac{1}{2}\,r_c^{2}-\,\lambda_{GB} \right) {\tfrac {\partial f}{\partial \lambda_{GB}}} + f^2-3\,f  \right) r_c^{D-5}\sqrt {f_0 }\,r_c^{5} \bigg]\\
\nonumber\\
\phi=\ &\frac { \left(D-2 \right)  \Omega_{D-3}\, r_c^{D-5}  \, \big( \lambda_{GB}\,(f-1) -\tfrac{1}{2}\,r_c^{2}\big) \tfrac{\partial f}{\partial q}}{8\pi \sqrt {f\,}}
\end{align}
\\
The uncharged case differs from the charged case only in the functional forms of $f$ and $f_0$ that appear in these expressions, as well as the lack of $\phi dQ$ term in the first law.

\bibliographystyle{JHEP} 
\bibliography{LBIB}

\end{document}